\documentclass[10pt,aps,prd,amsmath,amssymb,superscriptaddress,onecolumn,nofootinbib,showpacs,preprintnumbers,amsfonts, notitlepage]{revtex4-2}

\usepackage{amsmath,amssymb}
\usepackage{bm}
\usepackage{multirow}
\usepackage{xspace}
 \usepackage[usenames,dvipsnames]{xcolor}
\usepackage{colortbl}
\usepackage{upgreek}
\usepackage{tikz}
\usepackage{physics}

\usepackage{pifont}
\renewcommand{\arraystretch}{1.4}

\usepackage[colorlinks,hyperindex, 
           bookmarks=true,bookmarksopen=true]{hyperref}
          \hypersetup    {
        colorlinks=true,
       linkcolor=black,
        urlcolor=MidnightBlue,
   }

\usepackage[compat=1.1.0]{tikz-feynman}

\providecommand{\U}[1]{\protect\rule{.1in}{.1in}}
\newcommand{\gev}{\ensuremath{{\mathrm{\,Ge\kern -0.1em V}}}\xspace}
\newcommand{\gevsq}{\ensuremath{{\mathrm{\,Ge\kern -0.1em V}}^2}\xspace}
\newcommand{\mev}{\ensuremath{{\mathrm{\,Me\kern -0.1em V}}}\xspace}

\begin{document}

\title{Is $f_2(1950)$ the tensor glueball?}

\newcommand{\kielce}{Institute of Physics, Jan Kochanowski University, ul. Uniwersytecka 7, 25-406, Kielce, Poland}

\newcommand{\frankfurt}{Institute for Theoretical Physics, J. W. Goethe University,\\
Max-von-Laue-Straße. 1, 60438 Frankfurt am Main, Germany}

\author{Arthur Vereijken\footnote{
 \href{mailto:arthur.vereijken@gmail.com}{arthur.vereijken@gmail.com}}}
\affiliation{\kielce}

\author{Shahriyar Jafarzade\footnote{
\href{mailto:shahriyar.jzade@gmail.com}{shahriyar.jzade@gmail.com}}}
\affiliation{\kielce}

\author{Milena Piotrowska\footnote{
 \href{mailto:milena.piotrowska@ujk.edu.pl}{milena.piotrowska@ujk.edu.pl}} }
\affiliation{\kielce}

\author{Francesco Giacosa\footnote{
 \href{mailto:francesco.giacosa@gmail.com}{francesco.giacosa@gmail.com}}} 
\affiliation{\kielce}\affiliation{\frankfurt}

\begin{abstract}
\noindent
Glueballs remain an experimentally undiscovered expectation of QCD. Lattice QCD (As well as other theoretical approaches) predicts a spectrum of glueballs, with the tensor ($J^{PC}=2^{++}$) glueball being the second lightest, behind the scalar glueball. 
Here, using a chiral hadronic model, 
we compute decay ratios of the tensor glueball into various meson decay channels. We find the tensor glueball to primarily decay into two vector mesons, dominated by $\rho \rho $ and $K^*K^*$ channels. These results are compared to experimental data of decay rates of isoscalar tensor mesons. Based on this comparison, we make statements on the eligibility of these mesons as potential tensor glueball candidates: the resonance $f_2(1950)$ turns out to be, at present, the best match as being predominantly a tensor glueball. 
\end{abstract}

\maketitle

\section{Introduction}
\noindent
Glueballs are mesons made up by gluons only. They are one of the earliest predictions of Quantum Chromodynamics (QCD) that follow from the non-abelian nature of the $SU(3)$ gauge symmetry and confinement \cite{Gross:2022hyw}. 

The existence of glueballs has been since then confirmed by various theoretical approaches \cite{Mathieu:2008me,Crede:2008vw,Llanes-Estrada:2021evz,Chen:2022asf}, most notably Lattice QCD (LQCD), according to which a whole tower of such states has been computed \cite{Chen:2005mg,Morningstar:1999rf,Athenodorou:2020ani, Meyer:2004gx, Sakai:2022zdc,Chen:2021dvn,Bonanno:2022yjr} in the Yang-Mills (YM) sector of QCD (that is, QCD with without dynamical quarks; for an unquenched study see Ref. \cite{Gregory:2012hu}).
Similar outcomes were found in recent functional approaches in e.g. \cite{Huber:2021yfy,Pawlowski:2022zhh}.
The lightest gluonic state is a scalar ($J^{PC}=0^{++}$), the second-lightest a tensor ($J^{PC}=2^{++}$ (for the tensor/scalar mass ratio see\cite{Bennett:2020hqd}), and the third-lightest a pseudoscalar ($J^{PC}=0^{-+}$).
Quite interestingly, as shown in Ref. \cite{Trotti:2022knd}, the glueball spectrum of the recent LQCD compilation of Ref. \cite{Athenodorou:2020ani} generates a glueball resonance gas which is well consistent with the pressure of pure YM below $T_c$ as also evaluated in LQCD \cite{Borsanyi:2013bia}, thus being a further confirmation of the existence and accuracy of the LQCD masses.

Yet, although glueballs are such a long-standing prediction of QCD, their experimental status is still inconclusive. Admittedly, some notable candidates do exist, especially for the three lightest ones,  e.g. Refs. \cite{Llanes-Estrada:2021evz}). Nevertheless, the problem of a definitive identification of glueballs in experiments is made difficult by the mixing with nearby conventional quark-antiquark mesons and by their poorly known decay strength(s). 



In this work, we concentrate on the tensor glueball, which is is interesting because of various reasons. First, as already mentioned, is the second lightest gluonium. Second, the experimental observation of several isoscalar-tensor resonances such as $f_2(1430)$, $f_2(1565)$, $f_2(1640)$, $f_2(1810)$, $f_2(1910)$, $f_2(1950)$, $f_2(2010)$, $f_2(2150)$, $f_J(2220)$ etc. implies that one of them (especially close to 2 GeV, see below) could be the tensor glueball. The attempt to interpret one of those
$f_2$ states as a tensor glueball has a long history, see e.g \cite{Gross:2022hyw,Turnau:1984bb,Klempt:2000ud,Anisovich:2005xd}.
In particular the $f_J(2220)$ was historically considered as a good
candidate for the tensor glueball \cite{Anisovich:2004vj}, but, as we shall see, this is no longer the case.

In LQCD simulations various physical masses for the tensor glueball are predicted such as: 2150(30)(100) MeV \cite{Meyer:2004gx}, 2324(42)(32) MeV \cite{Sakai:2022zdc}, 2376(32)  MeV \cite{Athenodorou:2020ani}, 2390(30)(120) MeV \cite{Chen:2005mg}, 2400(25)(120) MeV \cite{Morningstar:1999rf}. 
Moreover, QCD sum rules predict the tensor glueball mass in the range of 2000-2300 MeV \cite{Narison:1996fm,Narison:2022paf}.  A recent functional  method result for the tensor glueball mass is around 2610 MeV \cite{Huber:2021yfy}. Holographic methods are also used to study glueballs for instance in 
\cite{Chen:2022goa,Brunner:2015yha,Brunner:2015oqa,Hechenberger:2023ljn, Brower:2000rp,Chen:2015zhh,Rinaldi:2021xat,Rinaldi:2022pgh,Elander:2018aub}, with tensor glueball masses ranging from 1900 to 4300 MeV depending on the model \cite{Chen:2015zhh,Rinaldi:2022pgh}. 
Thus, besides differences, the results converge on the existence of a tensor glueball in the region near 2000 MeV.  

In \cite{Brunner:2015oqa,Hechenberger:2023ljn} the decays of the tensor glueball are computed in the so-called Witten-Sakai-Sugimoto Model and it is found that this state is broad if the mass is above the $\rho \rho$ threshold. 
Other hadronic approaches are studied the decays of the tensor glueball \cite{Giacosa:2005bw,Li:2011xr}, where different decay ratios were presented. Yet, an explicit inclusion of chiral symmetry was then not considered.
The tensor glueball is also speculated to be related to the occurrence of OZI forbidden process such as $\pi^{-}p \rightarrow \phi \phi n$ \cite{Longacre:2004jn}.

A novel data analysis on $J/\psi$ decays is looking promising for the identification of the scalar and tensor glueballs \cite{Gross:2022hyw,Klempt:2022qjf,Sarantsev:2022tdi,Klempt:2022ipu}, with a tensor glueball mass of around 2210 MeV. Further recent measurable production of tensor glueballs in experiments are predicted in \cite{Kivel:2017tgt,Kou:2022ndm}.  

In view of this revival of both theoretical and experimental interest on the enigmatic tensor glueball, a quite natural question is what a chiral hadronic approach can tell about this state, especially in connection to certain decay ratios. 
The present paper deals exactly with this question: we study the tensor glueballs via a suitable extension of the Linear Sigma Model (LSM), called extended Linear Sigma model (eLSM), e.g. \cite{Parganlija:2012fy,Janowski:2014ppa}, in which chiral symmetry is linearly realized and undergoes both explicit and spontaneous symmetry breaking (SSB) and where, besides (pseudo)scalar mesons also (axial-)vector states and the dilaton are included from the very beginning. Previous applications of the eLSM to the scalar \cite{Janowski:2011gt,Janowski:2014ppa}, pseudoscalar \cite{Eshraim:2012jv}, and vector\cite{Giacosa:2016hrm} glueballs were performed. It is then natural to apply the formalism to the tensor gluonium.

Quite recently, we have employed the eLSM to study the tensor and axial-tensor mesons in Ref. \cite{Jafarzade:2022uqo}, which sets the formal framework to model the tensor glueball, as it has the same quantum numbers $J^{PC}=2^{++}$ as the tensor mesons (details of the fields and Lagrangians in Sec. \ref{sec2}). 
Within this framework, we evaluate (Sec. \ref{sec3}) various decay rates, most importantly the $\rho \rho$/$\pi \pi$ decay ratio, which turns out to be a prediction of the chiral approach (and its SSB). We then compare the outcomes to the isoscalar-tensor states in the range close to 2000 MeV and find out that the resonance $f_2(1950)$ is, according to present data, our best candidate. We also speculate (Sec. \ref{sec4}) about the partial decay widths of the glueball and the overall assignment of excited tensor states.  Finally, conclusions are outlined in Sec. \ref{sec5}.

\section{Chiral model and nonets}
\label{sec2}
\noindent
The eLSM is an effective model based on  chiral symmetry in the linear form, together with explicit and spontaneous breaking aimed to reproduce -at the hadronic level- basic QCD features. 

In this section we shall describe the fields entering the Lagrangian as well as the interaction terms. While the eLSM can contain many different fields and interactions, in this manuscript we shall only highlight those relevant for the decays of a tensor glueball. For a more complete review of the eLSM we refer to  \cite{Parganlija:2010fz,Parganlija:2012fy,Janowski:2014ppa,Eshraim:2020ucw} and references therein (applications at nonzero temperature and density can be found in e.g. Refs. \cite{Kovacs:2016juc,Kovacs:2022zcl,Heinz:2013hza}).
\subsection{Meson nonets}
The mesonic $\bar{q}q$ fields are contained in nonets. The list of decay products of the tensor glueball is, in matrix form, the following:
\begin{itemize}

    \item Vector mesons \{$\rho(770),$ $K^{\ast}(892),$ $\omega(782),$ $\phi(1020)$\} with the quantum number $J^{PC}=1^{--}$ ($^{3}S_{1}$)   can be classified to the following nonet
\begin{equation}
V^{\mu}=\frac{1}{\sqrt{2}}%
\begin{pmatrix}
\frac{\omega_{N}^{\mu}+\rho^{0\mu}}{\sqrt{2}} & \rho^{+\mu} &
K^{\ast+\mu}\\
\rho^{-\mu} & \frac{\omega_{N}^{\mu}-\rho^{0\mu}}{\sqrt{2}} & K^{\ast0\mu}\\
K^{\ast-\mu} & \bar{K}^{\ast0\mu} & \omega_{S}^{\mu}%
\end{pmatrix}
\,, \label{eq:vector_nonet}%
\end{equation}
where $\omega_{N}$ and $\omega_{S}$ are purely non-strange and strange,
respectively. The physical fields arise upon mixing
\begin{equation}
\left(
\begin{array}
[c]{c}%
\omega(782)\\
\phi(1020)
\end{array}
\right)  =\left(
\begin{array}
[c]{cc}%
\cos\beta_{V} & \sin\beta_{V}\\
-\sin\beta_{V} & \cos\beta_{V}%
\end{array}
\right)  \left(
\begin{array}
[c]{c}%
\omega_{N}\\
\omega_{S}%
\end{array}
\right)  \text{ ,}%
\end{equation}
where the small isoscalar-vector mixing angle $\beta_{V}=-3.9^{\circ}$ \cite{Workman:2022ynf}. The physical states $\phi$ and $\omega$
are predominantly strange and non-strange components, respectively. This is a reflection of the ``homochiral'' nature of these states \cite{Giacosa:2017pos}.

\item The chiral partners of the vector mesons \cite{Hatanaka:2008gu} having the quantum number $J^{PC}=1^{++}$ ($^{3}P_{1}$) forms the following nonet 
\begin{equation}
A_{1}^{\mu}=\frac{1}{\sqrt{2}}%
\begin{pmatrix}
\frac{f_{1,N}^{\mu}+a_{1}^{0\mu}}{\sqrt{2}} & a_{1}^{+\mu} & K_{1A}^{+\mu}\\
a_{1}^{-\mu} & \frac{f_{1,N}^{\mu}-a_{1}^{0\mu}}{\sqrt{2}} & K_{1A}^{0\mu}\\
K_{1A}^{-\mu} & \bar{K}_{1A}^{0\mu} & f_{1,S}^{\mu}%
\end{pmatrix}
\text{ }\,. \label{eq:avector_nonet}%
\end{equation}
The isoscalar sector reads
\begin{equation}
\left(
\begin{array}
[c]{c}%
f_{1}(1285)\\
f_{1}(1420)
\end{array}
\right)  =\left(
\begin{array}
[c]{cc}%
\cos\beta_{A_{1}} & \sin\beta_{A_{1}}\\
-\sin\beta_{A_{1}} & \cos\beta_{A_{1}}%
\end{array}
\right)  \left(
\begin{array}
[c]{c}%
f_{1,N}^{\mu}\\
f_{1,S}^{\mu}%
\end{array}
\right)  \text{ ,}%
\end{equation}
where the mixing angle $\beta_{A_{1}}$ is expected to be small because of the homochiral nature of the multiplet \cite{Giacosa:2017pos}, see also Ref. \cite{Divotgey:2013jba}. In the computations we shall set this angle to zero for simplicity.  
    \item Pseudoscalar mesons with $J^{PC}=0^{-+}$ ($^{1}S_{0}$) consisting of three pions, four kaons, the $\eta(547)$ and the $\eta^\prime(958)$. 
 They are collected into the following nonet with light-quark elements $P_{ij}\equiv2^{-1/2}\bar{q}_{j}i\gamma^{5}q_{i}$:
\begin{equation}
P=\frac{1}{\sqrt{2}}%
\begin{pmatrix}
\frac{\eta_{N}+\pi^{0}}{\sqrt{2}} & \pi^{+} & K^{+}\\
\pi^{-} & \frac{\eta_{N}-\pi^{0}}{\sqrt{2}} & K^{0}\\
K^{-} & \bar{K}^{0} & \eta_{S}%
\end{pmatrix}
\,\text{,}
\label{eq:pseudoscalar_nonet}%
\end{equation}
where $\eta_{N}\equiv\sqrt{1/2}\,(\bar{u}u+\bar{d}d)$ stands for the
purely non-strange state and $\eta_{S}\equiv\bar{s}s$ for the purely
strange one. 
The physical isoscalar fields appear upon mixing \ Ref. \cite{Kloe2} as:
\begin{equation}
\left(
\begin{array}
[c]{c}%
\eta(547)\\
\eta^{\prime}\equiv\eta(958)
\end{array}
\right)  =\left(
\begin{array}
[c]{cc}%
\cos\beta_{P} & \sin\beta_{P}\\
-\sin\beta_{P} & \cos\beta_{P}%
\end{array}
\right)  \left(
\begin{array}
[c]{c}%
\eta_{N}\\
\eta_{S}%
\end{array}
\right)  \text{ ,}
\end{equation}
with mixing angle $\beta_{P}=-43.4^{\circ}$. This sizable mixing angle is a consequence of the $U_{A}(1)$ axial anomaly \cite{Feldmann:1998vh,tHooft:1986ooh}. Namely, pseudoscalar mesons belong to a  ``heterochiral'' multiplet \cite{Giacosa:2017pos}.

\item The axial-vector matrix is shifted due to spontaneous chiral symmetry breaking. This breaking induces a mixing of pseudoscalar and axial-vector fields which allows the decay into pseudoscalar mesons. The shift is as follows:
\begin{align}
        A_{1\,\mu}\rightarrow A_{1\,\mu}+\partial_{\mu}\mathcal{P} ,\,\,  \mathcal{P} := \frac{1}{\sqrt{2}}
		\begin{pmatrix}
			\frac{Z_{\pi}w_{\pi}(\eta_N + \pi^0)}{\sqrt{2}} & Z_{\pi}w_{\pi}\pi^+ & Z_Kw_K K^+ \\
			Z_{\pi}w_{\pi}\pi^- & \frac{Z_{\pi}w_{\pi}(\eta_N - \pi^0)}{\sqrt{2}} &  Z_Kw_K K^0 \\
			 Z_Kw_K K^- &  Z_Kw_K \bar{K}^0 & Z_{\eta_S}w_{\eta_S}\eta_S
		\end{pmatrix}
		\label{shiftedP}
		\text{ ,}
    \end{align}

where the numerical values $Z_\pi=Z_{\eta_N}=1.709$, $Z_K=1.604$, $Z_{\eta_S}=1.539$ and $w_\pi=w_{\eta_N}=0.683\,\text{GeV}^{-1}$ , $w_K=0.611\, \text{GeV}^{-1}$ , $w_{\eta_S}=0.554 \,\text{GeV}^{-1}$ are taken from \cite{Parganlija:2012fy}. Note, this shift will be particularly important in this work since it allows to link different decay channels (such as $\rho \rho$/$\pi \pi$) that otherwise could not be connected by flavor symmetry alone. 
\item The $J^{PC}=2^{++}$ ($^{3}P_{2}$) tensor states with elements $T_{ij}^{\mu\nu}=2^{-1/2}\bar{q}
_{j}(i\gamma^{\mu}\partial^\nu+\cdots)q_{i}$: 
\begin{equation}
T^{\mu\nu}=\frac{1}{\sqrt{2}}%
\begin{pmatrix}
\frac{f_{2,N}^{\mu\nu}+a_{2}^{0\mu\nu}}{\sqrt{2}} & a_{2}^{+\mu\nu} &
K_{2}^{\ast+\mu\nu}\\
a_{2}^{-\mu\nu} & \frac{f_{2,N}^{\mu\nu}-a_{2}^{0\mu\nu}}{\sqrt{2}} &
K_{2}^{\ast0\mu\nu}\\
K_{2}^{\ast-\mu\nu} & \bar{K}_{2}^{\ast0\mu\nu} & f_{2,S}^{\mu\nu}%
\end{pmatrix}
\,. \label{eq:tensor_nonet}%
\end{equation}
The physical isoscalar-tensor states are
\begin{equation}\label{tens-mix}
\left(
\begin{array}
[c]{c}%
f_{2}(1270)\\
f_{2}^{\prime}(1525)
\end{array}
\right)  =\left(
\begin{array}
[c]{cc}%
\cos\beta_{T} & \sin\beta_{T}\\
-\sin\beta_{T} & \cos\beta_{T}%
\end{array}
\right)  \left(
\begin{array}
[c]{c}%
f_{2,N}\\
f_{2,S}%
\end{array}
\right)  \text{  ,}%
\end{equation}
where $\beta_{T} \simeq 5.7^{\circ}$ is the small mixing angle reported in the
PDG. Tensor mesons belong to a homochiral
multiplet, just as (axial-)vector states. We have no further experimental information about the chiral partner of this nonet $A_{2}^{\,\mu\nu}$ (see e.g.  \cite{Jafarzade:2022uqo} for recent phenomenological analyses of this nonet).
\end{itemize}
\subsection{Interaction terms}
\noindent
In this subsection, we list the chiral interaction involving a tensor glueball field $G_{2,\mu\nu}$.

A possible way to understand how the searched terms emerge is to consider the interaction terms that describe the decays of (axial-)tensor mesons in the recent Ref. \cite{Jafarzade:2022uqo}. What one needs to apply is the following replacement into the Lagrangians with ordinary tensor meson nonet:
\begin{equation}\label{G2-shift}
T_{\mu\nu}\longrightarrow \frac{1}{\sqrt{6}}G_{2,\mu\nu} \cdot\mathbf{1}_{3} 
\text{ ,}
\end{equation}
thus effectively realizing flavor blindness. Of course, the coupling constants must be renamed and are - at first - not known. Note, we follow the same convention that implies the normalization w.r.t. the flavor singlet mode.
Yet, the chirally invariant interaction terms of the tensor glueball with other mesons that we are going to introduce stand on their own and can be formally introduced without resorting to (axial-)tensor mesons.

The first term in the eLSM leading to tensor glueball decays involves solely left- and right-handed chiral fields: 
\begin{align}\label{lag1}
    \mathcal{L}_{\lambda}=\frac{\lambda}{\sqrt{6}} G_{2,\mu\nu}\Big(\text{Tr}\Big[ \{ L^{\mu}, L^{\nu}\}\Big]+\text{Tr}\Big[ \{R^{\mu}, R^{\nu}\} \Big]\Big)  
    \text{ ,}
\end{align}
where the chiral fields consist of the vector and axial-vector mesons which we shall define with nonets
\begin{align}
     L^{\mu}:= V^{\mu}+A_1^{\mu} \ \text{ , }  R^{\mu}:= V^{\mu}-A_1^{\mu} \text{ ,}
\end{align}
which transform as $L^{\mu} \to U_L L^{\mu}U_L^{\dagger} $, $ R^{\mu} \to U_{R}R^{\mu}U^{\dagger}_{R}$ under the chiral transformations of $U_L(3) \times U_R(3)$. The Lagrangian \eqref{lag1} leads to three kinematically allowed decay channels with the following expressions for the decay rates with three momentum
\begin{equation}
|\vec{k}_{a,b}|:=\frac{1}{2\,m_{G_2}}\sqrt{(m_{G_2}^{2}-m_{a}^{2}-m_{b}^{2}%
)^{2}-4m_{a}^{2}m_{b}^{2}}\ \text{ .} \label{eq:kf}%
\end{equation}
\begin{itemize}
    \item Decays of the tensor glueball into two vector mesons has the following decay rate formula
    \begin{align}\nonumber
\Gamma_{G_2\rightarrow V^{(1)}V^{(2)}}(m_{G_2},m_{v^{(1)}},m_{v^{(2)}})=\frac{\kappa_{gvv\,,i}\lambda^2|\vec{k}_{v^{(1)},v^{(2)}}|}{120\,\pi\,m_{G_2}^{2}}\Big(15 &+ \frac{5 |\vec{k}_{v^{(1)},v^{(2)}}|^2}{m_{v^{(1)}}^2}+\frac{5 |\vec{k}_{v^{(1)},v^{(2)}}|^2}{m_{v^{(2)}}^2}\\&+\frac{2|\vec{k}_{v^{(1)},v^{(2)}}|^4}{m_{v^{(1)}}^2m_{v^{(2)}}^2}\Big)\,\Theta
(m_{G_2}-m_{v^{(1)}}-m_{v^{(2)}})\,;
\label{eq:dec-gpv}
\end{align}
   
\item while into 2 pseudoscalar mesons (SSB-driven shift of Eq. \ref{shiftedP} applied twice)
 \begin{equation}
\Gamma_{G_2\longrightarrow P^{(1)}P^{(2)}}^{tl}(m_{G_2},m_{p^{(1)}},m_{p^{(2)}}%
)=\frac{\kappa_{gpp\,,i}\, \lambda^2\,|\vec{k}_{p^{(1)},p^{(2)}}|^{5}}{60\,\pi\,m_{G_2}^{2}}%
\Theta(m_{G_2}-m_{p^{(1)}}-m_{p^{(2)}})\,; \label{eq:dec-gpp}%
\end{equation}

\item and into the axial-vector and pseudoscalar mesons (SSB-driven shift of Eq. \ref{shiftedP} applied once)
\begin{equation}
\Gamma_{G_2\longrightarrow A_{1}P}^{tl}(m_{G_2},m_{a_{1}},m_{p})=\frac
{\kappa
_{gap\,,i}\,\lambda^2\,|\vec{k}_{a_{1},p}|^{3}}{120\,\pi\,m_{G_2}^{2}}
\left(5+\frac{2\,|\vec{k}_{a_{1},p}|^{2}}{m_{a_{1}}^{2}}\right)\,\Theta(m_{G_2}-m_{a_{1}}-m_{p})\,.
\label{eq:dec-gavp}
\end{equation}

\end{itemize}
The coupling constant $\lambda$ is not known and thus we are limited to computing branching ratios rather than decay rates. The (sometimes dimensionful) coefficients $\kappa_{g\circ\circ\,,i}$ are shown in table \ref{tabkgap}. For the 2-pseudoscalar channel, $\kappa$ has mass dimension of -4, in the vector-vector channel it is dimensionless, in the axial-vector plus pseudoscalar channel it has mass dimension -2, for the tensor and pseudoscalar channel it has mass dimension 2.
\\
\begin{table}[ptb]
\centering
\renewcommand{\arraystretch}{1.5}
\begin{tabular}
[c]{|l|c|}\hline
~~~~~~~~~Decay process & $\kappa
_{g\circ\circ\,,i}$\\\hline\hline
$\,\; G_{2} \longrightarrow\rho(770)\,\rho(770)$  & $ 1$\\\hline
$\,\; G_{2} \longrightarrow\overline{K}^\star(892)\,K^\star(892)$  & $ \frac{4}{3}$\\\hline
$\,\; G_{2} \longrightarrow\omega(782)\,\omega(782)$  & $ \frac{1}{3}$\\\hline
$\,\; G_{2} \longrightarrow\omega(782)\,\phi(1020)$  & $0$\\\hline
$\,\; G_{2} \longrightarrow\phi(1020)\,\phi(1020)$  & $ \frac{1}{3}$\\\hline\hline
$\,\; G_{2} \longrightarrow\pi\,\pi$  & $ (Z_\pi^2w_\pi^2)^2$\\\hline
$\,\; G_{2} \longrightarrow\bar{K}\,K$ & $\frac{4}{3}\times(Z_k^2w_k^2)^2$ \\\hline
$\,\; G_{2} \longrightarrow\eta\,\eta$   & $\frac{1}{3}\times (Z_{\eta_N}^2w_{\eta_N}^2\cos\beta_P^2+Z_{\eta_S}^2w_{\eta_S}^2\sin\beta_P^2)^2$\\\hline
$\,\; G_{2} \longrightarrow\eta\,\eta^{\prime}(958)$  & $\frac{1}{3}\times ( (Z_{\eta_N}^2w_{\eta_N}^2-Z_{\eta_S}^2w_{\eta_S}^2)\cos\beta_P\sin\beta_P)^2$\\\hline
$\,\;G_{2} \longrightarrow\eta^{\prime}(958)\,\eta^{\prime}(958)$ &  $\frac{1}{18}\times (Z_{\eta_S}^2w_{\eta_S}^2\cos\beta_P^2+Z_{\eta_N}^2w_{\eta_N}^2\sin\beta_P^2)^2$\\\hline\hline
$\,\; G_{2} \longrightarrow a_1(1260)\,\pi$ & $\frac{1}{2}\times (Z_\pi w_\pi)^2$\\\hline
$\,\; G_{2} \longrightarrow f_1(1285)\,\eta$ & $\frac{1}{6}(Z_{\eta_S}w_{\eta_S}\sin\beta_{A_1}\sin\beta_P+Z_{\eta_N}w_{\eta_N}\cos\beta_{A_1}\cos\beta_P)^{2}$\\\hline
$\,\; G_{2} \longrightarrow K_{1\,,A}\,K$ & $\frac{2}{3}\times(Z_k w_k)^2$\\\hline
$\,\;G_{2} \longrightarrow f_1(1420)\,\eta^{\prime}(958)$ & $\frac{1}{6}(Z_{\eta_N}w_{\eta_N}\sin\beta_{A_1}\sin\beta_P+Z_{\eta_S}w_{\eta_S}\cos\beta_{A_1}\cos\beta_P)^2$\\\hline
$\,\; G_{2} \longrightarrow f_1(1285)\,\eta^{\prime}(958)$ &
$\frac{1}{6}(Z_{\eta_S}w_{\eta_S}\sin\beta_{A_1}\cos\beta_P-Z_{\eta_N}w_{\eta_N}\cos\beta_{A_1}\sin\beta_P)^{2}$\\\hline
\hline
$\,\; G_{2} \longrightarrow f_{2}(1270) \,\eta$ & $\frac{1}{24}\big(\phi_N\cos{\beta_p\cos{\beta_T}}+\phi_S\sin{\beta_p\sin{\beta_T}}\big)^2$ \\
 \hline
$\,\; G_{2} \longrightarrow a_{2}(1320) \,\pi$ & $\frac{\phi^2_N}{8}$ \\
 \hline
 \hline
$\,\; G_{2} \longrightarrow K^*(892)\,K  + \text{c.c.} $& $\frac{1}{48} \left(\sqrt{2} \phi_N -2 \phi_S\right)^2 $ \\
\hline
\end{tabular}
\caption{Coefficients for the decay channel  $G_{2}\longrightarrow A+B$. $\phi_N \approx 0.158\, \text{GeV}$ and $\phi_S \approx 0.138\, \text{GeV}$ are due to the chiral condensate. The $\eta \eta^{\prime}$ coefficient is small because it only occurs due to the flavor symmetry breaking }%
\label{tabkgap}%
\end{table}
The second chirally invariant term we will use for tensor glueball decays is of the form
\begin{align}\label{lag2}
    \mathcal{L}_{\alpha}=\frac{\alpha}{\sqrt{6}} G_{2,\mu\nu}\Big(\text{Tr}\Big[ \Phi \textbf{R}^{\mu\nu}\Phi^\dagger\Big]+\text{Tr}\Big[ \Phi^\dagger\textbf{L}^{\mu\nu}\Phi\Big]\Big)  
    \text{ ,}
\end{align}
where $\Phi = S + i P$ is the linear combination of scalar\footnote{We do not study the decay to scalar mesons and so do not discuss the scalar nonet in this work.} and pseudoscalar nonets, and $ \mathbf{L}^{\mu\nu}=T^{\mu \nu}+A_2^{\mu \nu} ,\mathbf{R}^{\mu\nu}=T^{\mu \nu}-A_2^{\mu \nu}$ combine the tensor and axial-tensor nonets. Their chiral transformation rules are $\Phi \to U_L\Phi U^{\dagger}_R , \mathbf{R}^{\mu\nu} \to U_{R}\mathbf{R}^{\mu\nu}U^{\dagger}_{R} ,\mathbf{L}^{\mu\nu} \to U_{L}\mathbf{L}^{\mu\nu}U^{\dagger}_{L}$. The linear combination of the scalar and pseudoscalar contains the chiral condensate $\Phi = \Phi + \Phi_0 $ and so this term leads to the decay of tensor glueball into a tensor meson and pseudoscalar meson. The decay rate for this process is given by
\begin{align}
\Gamma_{G_2\longrightarrow TP}^{tl}(m_{G_2},m_{t},m_{p})=\frac{ \alpha^2|\vec{k}_{t,p}%
|}{2\,m_{G_2}^2\pi}\Big(1+\frac{4|\vec{k}_{t,p}%
|^4}{45m_{t}^4}+\frac{2|\vec{k}_{t,p}%
|^2}{3m_{t}^2}\Big)
\,\kappa_{g_2tp\,,i}\,
\Theta
(m_{G_2}-m_{t}-m_{p})\,.
\end{align}
The coupling $\alpha$ is not fixed, so branching ratios are only calculated for decays in this channel.
\\
\\
The third chiral Lagrangian describes the decay of a tensor glueball into a vector and pseudoscalar meson: 
\begin{align}\nonumber\label{chiralag-vp}
    \mathcal{L}_{c^{\text{ten}}}=\frac{c_1}{\sqrt{6}}\,\partial^{\mu}G_{2}^{\nu\alpha} \text{Tr} \Big[ \tilde{L}_{\mu\nu}\,\partial_{\alpha}\Phi\,\Phi^\dagger- \Phi^\dagger\,\partial_{\alpha}\Phi \tilde{R}_{\mu\nu}- \tilde{R}_{\mu\nu}\partial_{\alpha}\Phi^\dagger \Phi +\Phi \partial_{\alpha} \Phi^\dagger \tilde{L}_{\mu\nu}\Big]\\
   + \frac{c_2}{\sqrt{6}}\,\partial^{\mu}G_{2}^{\nu\alpha}\text{Tr} \Big[ \partial_{\alpha}\Phi \tilde{R}_{\mu\nu}\,\,\Phi^\dagger-\Phi^\dagger\,\tilde{L}_{\mu\nu}\partial_{\alpha}\Phi - \partial_{\alpha}\Phi^\dagger\tilde{L}_{\mu\nu} \Phi +\Phi\tilde{R}_{\mu\nu} \partial_{\alpha} \Phi^\dagger \Big]
   \text{ ,}
\end{align}
where $\tilde{L}_{\mu\nu}:=\frac{\varepsilon_{\mu\nu\rho\sigma}}{2}(\partial^{\rho}L^{\sigma}-\partial^{\sigma}L^{\rho})$ and similarly for $\tilde{R}_{\mu\nu}$.
 Defining $c:=c_1+c_2$, the tree-level decay rate formula reads
\begin{equation}
\Gamma_{G\longrightarrow VP}^{tl}(m_{G_2},m_{v},m_{p})=\frac{c^{\,2}\,|\vec{k}_{v,p}%
|^5}{40\,\pi}\,\kappa_{gvp\,,i}\,\Theta
(m_{G_2}-m_{v}-m_{p})\,,\label{eq:decaytvp}%
\end{equation}
The decay of $G_2$ into a vector and pseudoscalar is suppressed;namely, the only nonzero $\kappa$ is for $KK^*$ decay products, which is suppressed by a factor of $(\phi_N - \sqrt{2}\phi_S)$, and vanishes in the chiral limit. 

\section{Results for the decay ratios}
\label{sec3}
\begin{table}[ptb]
\centering
\renewcommand{\arraystretch}{2.} 
\begin{tabular}{|c|c|c|c|c|c|}
\hline
Branching Ratio   & theory & Branching Ratio   & theory &Branching Ratio   & theory \\ \hline\hline
	$ \frac{G_{2}(2210) \longrightarrow \bar{K}\, K}{G_{2}(2210) \longrightarrow  \pi\,\pi} $ & $0.4$ & $ \frac{G_{2}(2210) \longrightarrow \rho(770)\, \rho(770)}{G_{2}(2210) \longrightarrow  \pi\,\pi} $ & $55$ & $ \frac{G_{2}(2210)\longrightarrow a_1(1260)\,\pi}{G_{2}(2210) \longrightarrow  \pi\, \pi} $ & $0.24$ 
\\ \hline
 $ \frac{G_{2}(2210) \longrightarrow \eta \, \eta}{G_{2}(2210) \longrightarrow  \pi\,\pi}$  & $0.1$ & $ \frac{G_{2}(2210) \longrightarrow \bar{K}^\ast(892)\, \bar{K}^\ast(892)}{G_{2}(2210) \longrightarrow  \pi\,\pi} $ & $46$ & $ \frac{G_{2}(2210)\longrightarrow K_{1\,,A}\,K}{G_{2}(2210)\longrightarrow \pi\,\pi} $ & $0.08$
\\ \hline
$ \frac{G_{2}(2210) \longrightarrow \eta \, \eta^\prime}{G_{2}(2210) \longrightarrow  \pi\,\pi}$  & $0.004$ & $ \frac{G_{2}(2210) \longrightarrow \omega(782) \, \omega(782)}{G_{2}(2210) \longrightarrow  \pi\,\pi}$  & $18$ & $ \frac{G_{2}(2210)\longrightarrow f_1(1285)\,\eta}{G_{2}(2210)\longrightarrow \pi\,\pi}$  & $0.02$
\\ \hline
$ \frac{G_{2}(2210) \longrightarrow \eta^\prime \, \eta^\prime}{G_{2}(2210) \longrightarrow  \pi\,\pi}$  & $0.006$ &
$ \frac{G_{2}(2210) \longrightarrow \phi(1020) \,\phi(1020)}{G_{2}(2210) \longrightarrow  \pi\,\pi}$  & $6$ &	
$ \frac{G_{2}(2210)\longrightarrow f_1(1285)\,\eta}{G_{2}(2210)\longrightarrow \pi\,\pi}$  & $0.02$ 
\\ \hline
 & & & & $ \frac{G_{2}(2210)\longrightarrow f_1(1420)\,\eta}{G_{2}(2210)\longrightarrow \pi\,\pi}$  & $0.01$ \\ \hline
\end{tabular}%
\caption{Branching ratios of  $G_2$ w.r.t. $\pi \pi$. The columns are sorted as $PP$ on the left, $VV$ in the middle, and $AP$ on the right. The $VV$-$\pi\pi$ and $AP$-$\pi\pi$ ratios should be regarded as approximate due to large uncertainties (see text). }\label{tab:results}
\end{table}

\begin{figure*}[h]
       \centering
        \includegraphics[scale=0.53]{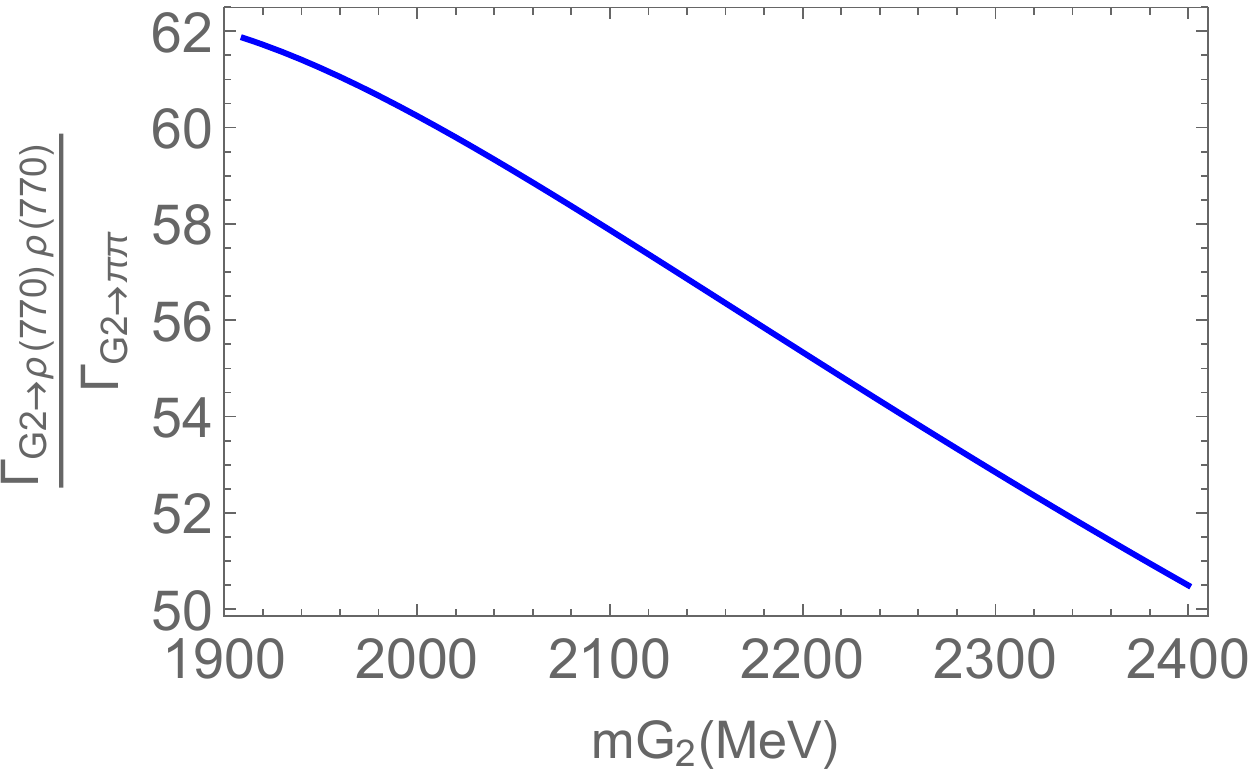}~~\includegraphics[scale=0.55]{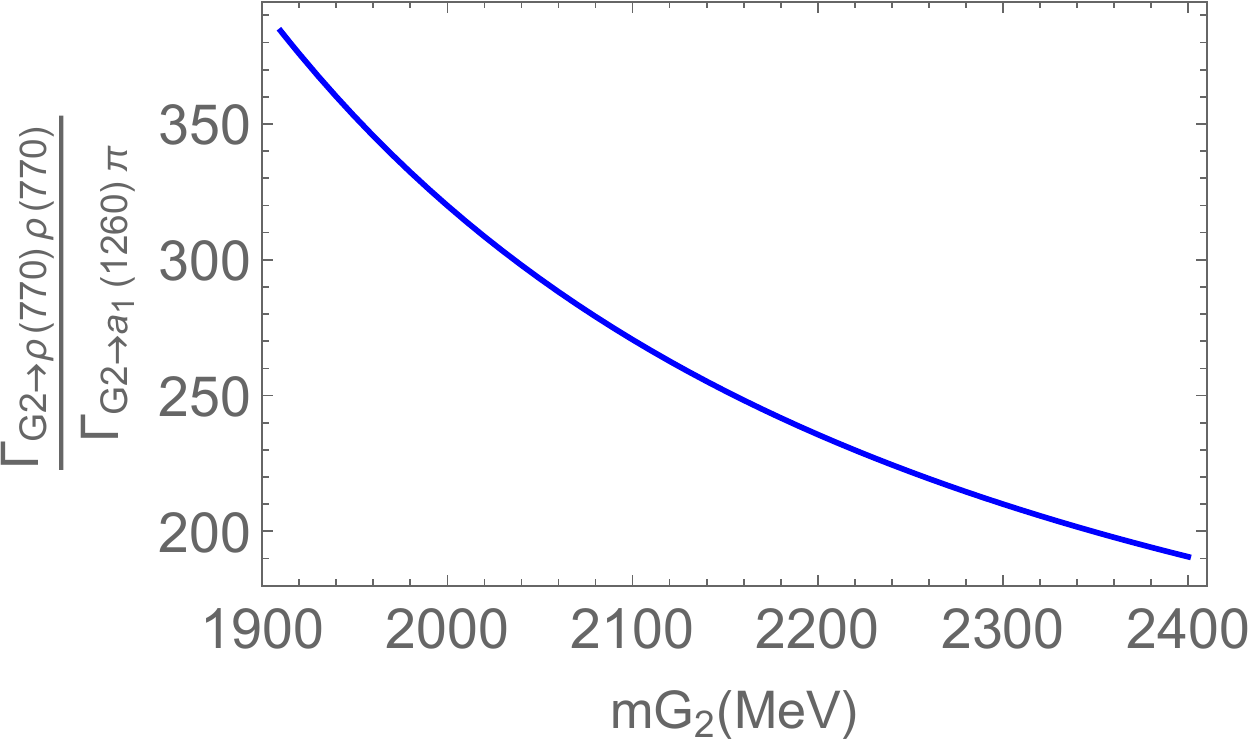}\\
        \caption{Estimates for the ratios $\rho \rho$/$\pi \pi$ (left) and $\rho \rho$/$a1(1260) \pi$ (right) as function of the tensor glueball mass.}
        \label{BR-plot}
   \end{figure*}
\noindent
Recent investigation devoted to the search of the tensor glueball in the BESIII data obtained an enhancement on the mass around 2210 MeV \cite{Klempt:2022qjf}. 
Thus, for illustrative purposes, we first present our decay ratios for this mass value.
\\
\noindent
Using the $\kappa
_{g\circ\circ\,,i}$ in Table \ref{tabkgap} and equations  \eqref{eq:dec-gpv}, \eqref{eq:dec-gpp}, \eqref{eq:dec-gavp}, we obtain the decay ratios shown in Table \ref{tab:results}. The decays are primarily to 2 vector mesons, with $\rho \rho$ and $K^*K^*$ being the two largest. The two $\rho$ mesons would further decay into four pions before reaching the detector in an experiment. Likewise the $K^*K^*$ pair decays to 2 $K \pi $ pairs. 

\begin{figure*}[h]
       \centering
       \includegraphics[scale=0.95]{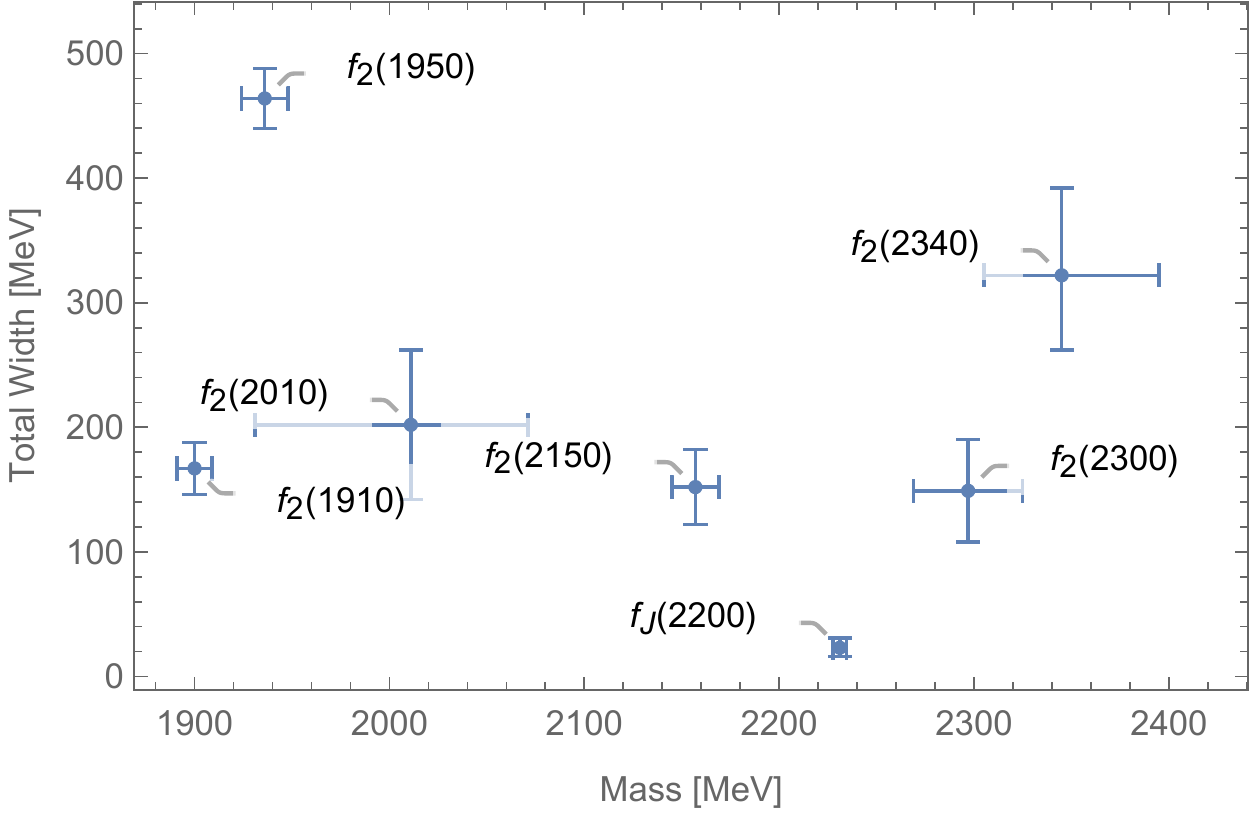}\\
        \caption{Masses and widths of isoscalar-tensor resonances heavier than 1.9 GeV \cite{Workman:2022ynf}. }
        \label{list}
   \end{figure*}

It should be stressed that we expect a quite large indeterminacy of our results, in particular in connection to the $\rho \rho$-$\pi \pi$ ratio, that is at least of 50 percent. In fact, as discussed in Ref. \cite{Jafarzade:2022uqo}, the $\rho \rho$ mode involves (because of the SSB-shift) the factor $Z_{\pi}^4$, thus even a quite small indeterminacy of  $Z_{\pi}$ may generate large changes. The error on $Z_{\pi}$ is of the order of 20\% , that translates into a factor two for $\rho \rho$. 
Yet, the main point of our study is that the ratio $\rho \rho$-$\pi \pi$ is expected to be large, as shown in figure \ref{BR-plot}.
A similar dominance of $\rho \rho$ and $K^*K^*$ decay products was found in \cite{Brunner:2015oqa} in a holographic model, but to a somewhat lesser extent (with a ratio of around 10, depending on the parameters). 
\\
Potential glueball candidates and some relevant data is given in table \ref{tab:PDGdata} and in Fig.\ref{list}. While the mass value in table \ref{tab:results} is not in line with every glueball candidate, it gives an overall picture on the decay ratios.  One should also keep in mind that lattice determinations still have sizable errors (of the order of 100-200 MeV) and do not include the role of meson-meson loops, that can be quite relevant if the tensor glueball turns out to be broad (this being our favored scenario).


\begin{table}[ptb]
\centering
\renewcommand{\arraystretch}{2.} 
\begin{tabular}{|c|c|c|c|}
\hline
Resonances & Masses (MeV) & Decay Widths (MeV) & Decay Channels   \\ \hline
$f_2(1910)$   & $1900\pm 9$       &  $167\pm 21$            &   $\pi\pi$, $KK$, $\eta\eta$, $\omega\omega$, $\eta\eta^\prime$, $\eta^\prime\eta^\prime$, $\rho\rho$, $a_{2}(1320)\pi$, $f_{2}(1270)\eta$                                                                       \\ \hline
$f_2(1950)$   &  $1936\pm 12$      &  $464\pm 24$            &   $\pi\pi$,   $KK$, $\eta\eta$, $K^{*}K^{*}$, $4\pi$           \\ \hline
$f_2(2010) $  & $2011^{+60}_{-80}$       &  $202\pm 60$            & $KK$, $\phi\phi$            \\ \hline
$f_2(2150)$   & $2157\pm 12$        &  $152\pm 30$            &  $\pi\pi$, $\eta\eta$, $KK$, $a_{2}(1320)\pi$, $f_{2}(1270)\eta$   \\ \hline
$f_J(2220)$   &   $2231.1\pm 3.5 $     &   $23^{+8}_{-7}$           & $\eta\eta^\prime$               \\ \hline
$f_2(2300)$   &  $2297\pm 28$        &    $149\pm 41$           &     $KK$,    $ \phi\phi$        \\ \hline
$f_2(2340)$   &  $2345^{+50}_{-40}$      &  $322^{+70}_{-60}$            & $\eta\eta$, $\phi\phi$              \\ \hline
\end{tabular}
\caption{Spin-2 resonances heavier than 1.9 GeV listed in PDG \cite{Workman:2022ynf}.}
\label{tab:PDGdata}
\end{table}

\begin{table}[h]
\renewcommand{\arraystretch}{2.} 
\begin{tabular}{|c|c|c|c|}
\hline
Resonances & Branching Ratios       & PDG                   & Model Prediction 
\\ \hline \hline
$f_2(1910)$   & $\rho(770)\rho(770) / \omega(782)\omega(782)$ &   $2.6\pm 0.4$ &    $3.1$        \\ 
\cline{1-4}
 \hline
$f_2(1910)$   & $f_2(1270)\eta/a_2(1320)\pi$ &   $0.09\pm 0.05$ &    $0.07$       \\ 

\cline{1-4}
 \hline
$f_2(1910)$   & $\eta \eta / \eta\eta^\prime(958)$ &   $<0.05$ &    $\sim 8$       \\ 

\cline{1-4}
 \hline
$f_2(1910)$   & $\omega(782) \omega(782) /\eta \eta\prime(958)$ &   $2.6 \pm 0.6$  &    $\sim 200$       \\ 
\hline \hline
\cline{1-4}
$f_2(1950)$   & $\eta\eta/\pi\pi$ &     $0.14 \pm 0.05$                  & $0.081$                       \\ 
\cline{1-4}
\hline
$f_2(1950)$   & $K\overline{K}/\pi\pi$ &     $\sim 0.8$                  & $0.32$                       \\ 
\cline{1-4}
\hline
$f_2(1950)$   & $4\pi/\eta\eta$ &     $>200$                  & $>700$                       \\ 

\hline \hline
\cline{1-4}
\hline
$f_2(2150)$   & $f_2(1270)\eta/a_2(1320)\pi$ &     $0.79\pm 0.11$                  & $0.1$                       \\ 
\cline{1-4}
\hline
$f_2(2150)$   & $K\overline{K} / \eta \eta$ &     $1.28\pm 0.23$                 & $\sim 4$                       \\ 
\cline{1-4}
\hline
$f_2(2150)$   & $\pi \pi / \eta \eta$ &     $<0.33$                 & $\sim 10$                       \\ 
\cline{1-4}
\hline \hline
$f_J(2220)$   & $\pi \pi / K\overline{K}$ &     $1.0 \pm 0.5$                 & $\sim 2.5$                       \\ 

\hline
\end{tabular}
\caption{Decay ratios for the decay channels with available data.}
\label{BR-table}
\end{table}


Below we discuss each tensor-glueball candidate separately. The experimental data and theoretical predictions used for this discussion are shown in table \ref{BR-table}.

\begin{itemize}
    \item The meson $f_2(1910)$ has a width of $167\pm 21$ MeV and it decays into (among others) $\eta \eta$ and $K\overline{K}$. The decay ratio $\rho(770)\rho(770) / \omega(782)\omega(782)$ of about $2.6\pm 0.4 $ is not far from the theoretical result of $3.1$, and the data on the decay ratio of $f_2(1270)\eta/a_2(1320)\pi$ also agrees with the theory. Yet, this state cannot be mainly gluonic since the experimental ratio $\eta \eta / \eta\eta^\prime(958)$ is less than $0.05$, while the theoretical result is much larger (about $8$), and the ratio of $\omega \omega $ to $\eta \eta\prime$ is very large in theory, but only $2.6 \pm 0.6$ in data. Summarizing, the $\eta\eta^\prime(958)$ mode is a clear drawback for $f_2(1910)$ being predominantly gluonic. 
\item The meson $f_2(1950)$ decays into  $\eta \eta$, $\pi \pi$, 
$K \bar{K}$, as well as $K^{*}K^{*}$ and $4\pi$ modes, the latter likely to emerge from $\rho \rho$. The experimental ratio $\eta \eta / \pi \pi$ of $0.14 \pm 0.05$ agrees with theory, as well as the $K \bar{K} / \pi \pi$. Both in experiment and theory we have only lower bounds of the ratio $4\pi /\eta\eta$ - where we assume $\rho \rho$ decays into 4 pions - but both agree that this ratio is large.
While the theory fits the data well, its large total decay width of $460$ MeV would imply a broad tensor glueball candidate. 
Quite interestingly, the decay 
$J/\psi\rightarrow\gamma K^{\ast}(892)\bar{K}^{\ast}(892)$ shows a relatively large branching ratio of 
$\left(
7.0\pm2.2\right)  10^{-4}$. A strong coupling to  $J/\psi$ is an expected feature of a tensor glueball.

\item The resonance $f_2(2010)$ has a total decay width of $202 \pm 60$ MeV. Yet, only $K\overline{K}$ and $\phi(1020)\phi(1020)$ decays have been seen. This fact suggests a large strange-antistrange content for this resonance, rather than a predominantly gluonic state, see also the discussions in Sec. IV. 
\item In view of the LQCD prediction for the tensor glueball mass around 2.2 GeV, one of the closest resonances is $f_2(2150)$. Yet, the ratio $K\overline{K} / \eta \eta$  is $1.28\pm 0.23$, while the theoretical prediction is about $4$. Similarly, the ratio  of $\pi \pi / \eta \eta$ is experimentally less than 
$ 0.33$, while the theoretical estimate is about $10$. The predicted ratio of 0.1 for $f_2(1270)\eta/a_2(1320)\pi$ also does not fit the data of $0.79 \pm 0.11$
\item The meson $f_J(2220)$ (with $J= 2$ or $4$) was historically treated as a good candidate \cite{Anisovich:2005xd}. However, in light of the new PDG compilation, most of the decays that the theory predicts are not seen experimentally. Moreover, the only channel denoted as `seen' is the $\eta \eta'(958)$ mode, which is expected to be extremely suppressed for a glueball state. PDG data offers one the decay ratio for $ \pi \pi / K \bar{K}$ of $ 1.0 \pm 0.5$ even though in the `decay modes' table these channels are marked as not seen) while the theoretical prediction is about 2.5. For these reasons, we conclude that $f_J(2220)$ should not be considered as a good candidate any longer.  
\item The resonance $f_2(2300)$, with a total width of $149\pm 41$ MeV, decays only into $K\overline{K}$ and $\phi \phi$, thus suggesting that it is predominantly a strange-antistrange object. 
\item Finally, the resonance $f_2(2340)$ decays into $\eta \eta$ and $\phi(1020) \phi(1020)$ that may also imply a large strange-antistrange component.
Yet, both latter resonances should be investigated in more details in the future, especially for what concerns other possible decay modes of these resonances. 
We also refer to the discussion in the next Section for additional discussions. 
\end{itemize}

From the considerations above, it turns out that the resonances $f_{2}(1950)$ seems to be the best fit, although this would imply a broad tensor glueball state. Namely, all other states seem disfavored for various reasons, such as specific branching or decay ratios with available data, ($f_{2}(1910)$ and the $f_{2}(2150)$) or decays in strange states only ($f_{2}(2010), f_{2}(2300),$ and $f_{2}(2340)$).

It is of primary importance to monitor the experimental status of the states above in the future. In particular, the analysis for the states $f_{J}(2220), f_{2}(2300),$ and $f_{2}(2340)$ would benefit from more experimental data, with special attention to the latter broad one.


\section{Discussions}
\label{sec4}
In this section, we discuss two additional important points. The first one
addresses the actual partial decay widths of the tensor glueball; while a rigorous treatment is not possible within our framework, a `guess' is achieved
by using large-$N_{c}$ arguments and the decays of the conventional
ground-state tensor mesons. The second point discusses the assignment of
various tensor states as radially and orbitally excited conventional tensor
mesons. In this framework, the tensor glueball should be a `supernumerary'
state that does not fit into the quark-antiquark nonet picture.

For the first point, let us consider the conventional mesons $f_{2}\equiv
f_{2}(1270)\simeq\sqrt{1/2}(\bar{u}u+\bar{d}d)$ and $f_{2}^{\prime}\equiv
f_{2}^{\prime}(1525)\simeq\bar{s}s$, whose decays into $\pi\pi$ are well
known: $\Gamma_{f_{2}\rightarrow\pi\pi}=157.2$ MeV and $\Gamma_{f_{2}^{\prime
}\rightarrow\pi\pi}=0.71$ MeV (for our qualitative purposes, we neglect the
anyhow small errors). The amplitude for the decay $A_{f_{2}\rightarrow
\pi\pi}$ requires the creation of a single $\bar{q}q$ pair from the vacuum and
scales as $1/\sqrt{N_{c}},$ where $N_{c}$ is the number of colors. On the other
hand, the amplitude $A_{f_{2}^{\prime}\rightarrow\pi\pi}$ implies that
$\bar{s}s$ first converts into two gluons ($gg$) that subsequently transforms
into $\sqrt{1/2}(\bar{u}u+\bar{d}d)$ (the very same mechanism is responsible for
a small (about $3^{\circ}$) but nonzero mixing angle of the physical states in
the strange-nonstrange basis \cite{Jafarzade:2022uqo}).
Schematically:
\begin{equation}
\bar{s}s\rightarrow gg\rightarrow\sqrt{1/2}(\bar{u}u+\bar{d}d)\text{ .}%
\end{equation}
The amplitude $A_{f_{2}^{\prime}\rightarrow\pi\pi}$ scales as $1/N_{c}^{3/2}$
and is OZI \cite{OKUBO1963165,Zweig:1964jf,Iizuka:1966wu} suppressed w.r.t. the previous one. In order to be more specific,
let us consider the transition Hamiltonian $H_{int}=\lambda\left(  \left\vert
\bar{u}u\right\rangle \left\langle gg\right\vert +\left\vert \bar
{d}d\right\rangle \left\langle gg\right\vert +\left\vert \bar{s}s\right\rangle
\left\langle gg\right\vert +h.c.\right)  $, where $\lambda$ controls the
mixing and therefore scales as $1/\sqrt{N_{c}}.$ Then: $A_{f_{2}^{\prime
}\rightarrow\pi\pi}\simeq\sqrt{2}\lambda^{2}A_{f_{2}\rightarrow\pi\pi},$
hence $\Gamma_{f_{2}^{\prime}\rightarrow\pi\pi}\simeq2\lambda^{4}\Gamma
_{f_{2}\rightarrow\pi\pi},$ implying $\lambda\simeq0.22$. 

Next, let us consider the tensor glueball decay into $\pi\pi$. Intuitively
speaking, it is at an `intermediate stage', since it starts with a $gg$ pair.
One has: $A_{G_{2}\rightarrow\pi\pi}\simeq\sqrt{2}\lambda A_{f_{2}%
\rightarrow\pi\pi},$ then $\Gamma_{G_{2}\rightarrow\pi\pi}\simeq2\lambda
^{2}\Gamma_{f_{2}\rightarrow\pi\pi}\simeq\sqrt{2}\sqrt{\Gamma_{f_{2}%
\rightarrow\pi\pi}\Gamma_{f_{2}^{\prime}\rightarrow\pi\pi}}\simeq15$ MeV.
(Note, for a similar idea for the estimate of the coupling of glueballs to
mesons, see Refs. \cite{Toki:1996si,Giacosa:2017eqy}.) $\Gamma_{G_{2}%
\rightarrow\pi\pi}$ scales as $1/N_{c}^{2}$ (as expected for glueballs), thus
realizing the expectation $\Gamma_{f_{2}^{\prime}\rightarrow\pi\pi}%
<\Gamma_{G_{2}\rightarrow\pi\pi}<\Gamma_{f_{2}\rightarrow\pi\pi}.$

Of course, the estimate $\Gamma_{G_{2}\rightarrow\pi\pi}\simeq15$ MeV is only
a very rough approximation for various reasons: it does not take into account
phase space (that would increase the glueball width) or form factors (that
would decrease it). It also avoids a microscopic evaluation\ (which is a
difficult task). Yet, it gives an idea on how large the $\pi\pi$ mode (and as
a consequence other $PP$ channels) could be, see Table \ref{gluebalesti}.

\begin{table}[pt]
\centering%
\begin{tabular}
[c]{|c|c|}\hline
~~~~~~~~~Decay process & $\Gamma_{i}(\text{MeV})$\\\hline\hline
$\,\;G_{2}\rightarrow\pi\,\pi$ & $\sim15$\\\hline
$\,\;G_{2}\rightarrow\bar{K}\,K$ & $\sim6$\\\hline
$\,\;G_{2}\rightarrow\eta\,\eta$ & $\sim1.6$\\\hline
$\,\;G_{2}\rightarrow\eta\,\eta^{\prime}(958)$ & $\sim0.06$\\\hline
$\,\;G_{2}\rightarrow\eta^{\prime}(958)\,\eta^{\prime}(958)$ &
$\sim0.08$\\\hline
\end{tabular}
\caption{Estimations of the decay channel $G_{2}\rightarrow
PP$ . }%
\label{gluebalesti}%
\end{table}

It is interesting to point out that our $\pi\pi$ decay rate is comparable to the one obtained within so-called
Witten-Sakai-Sugimoto model \cite{Brunner:2015oqa}, in which $\Gamma
_{G_{2}\rightarrow\pi\pi}/m_{G(2000)}\simeq0.014,$ thus implying
$\Gamma_{G_{2}\rightarrow\pi\pi}\simeq28$ MeV.

As a consequence of such decay width into $\pi\pi,$ a large decay width into
$\rho\rho\rightarrow\pi\pi\pi\pi$ is expected due to the evaluated large
$\rho\rho/\pi\pi$ ratio.

Next, the second point to discuss relates to the assignment of quark-antiquark tensor
states. Namely, up to now, we have considered all isoscalar $f_{2}$ states in
the energy region about 2 GeV `democratically' as putative tensor glueball
candidates. Yet, it is clear that many (if not all) of them should be rather
interpreted as standard $\bar{q}q$ objects.

For this reason, it is useful to classify them accordingly. While the
ground-state tensor mesons (with spectroscopic notation $1^{3}P_{2}$ are
well established as {$a_{2}(1320)$, $K_{2}^{\ast}(1430)$, $f_{2}%
(1270)$, and $f_{2}^{\prime}(1525)$}, one expects a nonet of radially excited
tensor states with $2^{3}P_{2}$ as well as a nonet of orbitally excited tensor
states with $1^{3}F_{2}$. The former are predicted to be lighter than the
latter \cite{isgur1985}, what is in agreement with the excited vector mesons
\cite{Piotrowska:2017rgt}.

The nonet of radially excited tensor mesons contains the isotriplet
$a_{2}(1700)$, the isodoublet states $K_{2}^{\ast}(1980)$, the isoscalar
(mostly nonstrange) $f_{2}(1640)$. The $\bar{s}s$ member of the nonet may be
assigned to $f_{2}(1910)$, or $f_{2}(1950)$, or $f_{2}(2010)$.

Indeed, the quark model review of the PDG \cite{Workman:2022ynf} considers $f_{2}(1950)$ as a
possible $\bar{s}s$ state. However, the decay of that state does not fit quite
well with a predominantly $\bar{s}s$ object. For instance, the experimental
$\pi\pi/KK$ ratio is of order unity, while a $\bar{s}s$ state should imply a
small ratio. In fact, the radially excited tensor states should display a
small isoscalar mixing angle, just as the ground state (see discussion above).
Moreover, the mass is too close to the tensor member $K_{2}^{\ast}(1980)$. Note that, resonance ``$f_2(2000)$'' included in the further states list of PDG, is proposed as a tensor glueball candidate in Ref.\cite{Anisovich:2005xd} due to its poorly fitted location in the Regge trajectory of spin-2 states. Similar to $f_{2}(1950)$, it has also a broad decay width.

The state $f_{2}(1910)$ is presently omitted from the summary table and it is
not clear if it corresponds to an independent pole (for instance, the PDG
notes that the $\bar{K}K$ mode could be well correspond to $f_{2}(1950)$). Its
mass is even lighter than $K_{2}^{\ast}(1980)$, what disfavors this state as
being an $\bar{s}s$. The decays -which are quite uncertain- confirm this view:
the modes $\rho\rho$, and $a_{2}(1320)\pi$ should be suppressed, and the
latter should be smaller then $f_{2}(1270)\eta$, contrary to data (see above).
On the other hand, the state $f_{2}(2010)$ is well established and
decays into $\phi\phi$ and $KK$, which indicates a strange content.

Summarizing the discussion above, we temptatively identify the nonet of
radially excited tensor mesons as
\begin{equation}
{a_{2}(1700),\ K_{2}^{\ast}(1980),\ f_{2}(1640),\ f_{2}(2010)}\text{ with
}2^{3}P_{2}\text{ and }J^{PC}=2^{++}.
\end{equation}
Within this assignment and upon neglecting the unsettled $f_{2}(1910)$, the
state $f_{2}(1950)$ may be seen as supernumerary. This argumentation can be an
additional hint toward the exotic nature of $f_{2}(1950)$.

Next, what about the orbitally excited states? Here the situation is much more
unclear. There are no isotriplet or isodoublet states that could be used to
identify the nonet. In the listing of the PDG one has $f_{2}(2150)$ (status
unclear, omitted from the summary table) and $f_{2}(2300)$ and $f_{2}(2340)$.
The latter two states have a prominent decay into $\phi\phi$, then, due to the
vicinity of mass, one may regard them as a unique state corresponding to
$\bar{s}s$ resonance. On the other hand, $f_{2}(2150)$ could be tentatively
correspond to a nonstrange isoscalar state belonging to the next radially
excited nonet $3^{3}P_{2}$. Definitely, more data and studies are needed for
these excited tensor states.

\section{Conclusions}
\label{sec5}

\begin{table}[ptb]
\centering
\renewcommand{\arraystretch}{2.} 
\begin{tabular}{|c|c|}
\hline
Resonances & Interpretation status   \\ \hline
$f_2(1910)$   &  Agreement with some data, but excluded by  $\eta \eta / \eta\eta^\prime$ and $\omega \omega/\eta \eta\prime$ ratios                       \\  \hline
$f_2(1950)$   &   $\eta \eta / \pi \pi$ agrees with data, no contradictions with data, but implies broad tensor glueball 
\\
& Best fit as predominantly glueball among considered resonances
\\ \hline
$f_2(2010)$   &  Likely primarily strange-antistrange content           \\ \hline
$f_2(2150)$   & All available data contradicts theoretical prediction   \\ \hline
$f_J(2220)$   & Data on $ \pi \pi / K \bar{K}$ disagrees with theory, largest predicted decay channels are not seen              \\ \hline
$f_2(2300)$   &        Likely primarily strange-antistrange content     \\ \hline
$f_2(2340)$   &    Likely primarily strange-antistrange content, would also imply a broad glueball          \\ \hline
\end{tabular}
\caption{Spin 2 resonances and the status as the tensor glueball.}
\label{tab:conclusion}
\end{table}

\noindent
In this work, we have applied the eLSM to the study of the tensor glueball, constructing chirally invariant Lagrangians describing the tensor glueball decays. From this we computed tensor glueball decay ratios, with dominant decay channels being the vector-vector decay products, especially $\rho \rho$ and $K^{*}\bar{K}^{*}$. 
A quite large tensor glueball follows from the large predicted $\rho \rho/\pi\pi$ ratio, with the $\pi \pi$ mode being of the order of 10 MeV.
Moreover, we also predict a very small decay width of the tensor glueball into $K^*(892)K$, which render this mode potentially interesting to exclude eventual glueball candidates in the future.
\\
We compared the theoretical predictions to the experimental data. At present, the best match is for the resonance  $f_{2}(1950)$, implying that the tensor glueball is a relatively broad state. The $f_{2}(1950)$ might be thought of as too light to be the tensor glueball, which - according to most lattice studies - has its mass in the range 2.2-2.4 GeV. However, unquenching effects included in additional meson-meson loop corrections are expected to bring the glueball mass down. The large width $f_{2}(1950)$ of the glueball is indeed in agreement with this view.
\\
Here, the tensor glueball mixing with other conventional meson states was not taken into account. While mixing with the ground state tensor meson is expected to be small due to the large mass difference (recently, a small mixing of the pseudoscalar glueball and the $\eta$ was studied on the lattice in \cite{Jiang:2022ffl}), this could be not the case for the nearby excited tensor states. The generalization to the eLSM is in principle possible and can be undertaken once better data, both form experiments and from lattice, will be available. 

In the future, more information for decays of all tensor states into vector-vector, pseudoscalar-pseudoscalar as well as ground-state tensor-pseudoscalar would be very helpful to constrain models and falsify different scenarios. Moreover, also the decay of $J/\psi$ into tensors as well as radiative (such as photon-photon) decays of tensor states could be of great use. In particular, more information about the broad state $f_2(2340)$ could shed light on its nature.

Another interesting future line of research is the study of glueball molecules \cite{Giacosa:2021brl,Petrov:2022rkn}. While two scalar glueballs may interact and form a bound state (which is stable in pure YM), the question for the analogous tensor-tensor case (and also tensor-scalar and heavier glueballs, such as the pseudoscalar one) is unsettled.

\section*{Acknowledgements}
A.V. and F.G. acknowledge financial support from the Polish National Science Centre (NCN) via the OPUS project 2019/ 33/B/ST2/00613. S.J.  acknowledges the financial support through the project ``Development Accelerator of the Jan Kochanowski University of Kielce'', co-financed by the European Union under the European Social Fund, with no. POWR.03.05. 00-00-Z212 / 18.

\bibliography{tensor-glueball}
\end{document}